\documentstyle [aps,prd,epsf,floats]{revtex}
\tighten
\begin{document}
\title{Lattice study of classical inflaton decay}

\author{Tomislav Prokopec\footnote{E-mail: tomislav@mail.lns.cornell.edu}
 and Thomas G. Roos\footnote{E-mail: roost@mail.lns.cornell.edu}}
\address{Newman Laboratory of Nuclear Studies, 
         Cornell University, Ithaca, NY 14853}

\maketitle

\begin{abstract}
We study numerically the decay of the inflaton by solving the full 
non-linear equations of motion on the lattice. We confirm that parametric
resonance is effective in transferring energy from the inflaton to a
scalar field as long as the self-interactions of the second field are
very small. However, in the very broad resonance case ($q\gg 1$)
the decay rate is limited by scatterings, which significantly
slows down the decay. We also find that the inflaton
cannot decay via parametric resonance into a scalar field with
moderate self-interactions. This means that
the preheating stage may be completely absent in many natural
inflationary models.
\end{abstract}


\section{Introduction}
According to inflationary scenarios of cosmic evolution the Universe 
undergoes a period of exponential expansion after which it is 
essentially devoid of matter. At this stage the energy density is
almost entirely in the large oscillating expectation value
(EV) of the scalar field that drives inflation, the inflaton. The 
theory of reheating is concerned with the question how this energy is 
transferred from the inflaton to other fields and eventually 
thermalized. Early calculations assumed that the inflaton would decay
to lighter particles via perturbative decay 
\cite{StarobinskiiDolgovLindeAbbottFarhiWise}. 
It was only recently realized \cite{TraschenBrandenberger} 
that non-perturbative 
mechanisms may completely dominate the reheating process. In
particular the classical phenomenon of parametric resonance may lead
to explosive particle production and very rapid decay of the inflaton
field \cite{KLS}. 
This stage of the evolution, called {\it preheating} in the
literature, leads to very different physics than would be obtained
from slow perturbative decay. Different aspects have been investigated
by many authors \cite{allres}.
 
In this paper we present the first fully non-linear study of inflaton
decay into a second scalar field. By fully non-linear we mean that we
take into account the back reaction of the created particles on the
equations of motion as well as their scattering. This is accomplished
by integrating the classical equations of motion of the system with
certain random initial conditions, a technique that was first applied
to the one field case in \cite{TK}. The system is
modeled by a simple two scalar theory with effective potential
\begin{equation}
V(\phi,\chi) = \frac{1}{2} m^2\phi^{2} + \frac{1}{2}\mu^2\chi^{2} 
+\frac{\lambda_{\phi}}{4}\phi^{4} + 
\frac{\lambda_{\chi}}{4}\chi^{4} + \frac{g}{2} 
\phi^{2}\chi^{2}. 
\end{equation} 
We will focus on two general classes of inflaton potentials $V(\phi)$: 
type I for which $\phi \ll m/\sqrt {\lambda_\phi}$ (for these we will set
$\lambda_\phi=0$), and type II for which 
$\phi \gg m/\sqrt \lambda_\phi$ (for
these we will set $m=0$). In both cases we will assume that there is a 
channel for the inflaton to decay into a light scalar and thus set
the tree level mass $\mu$ of the $\chi$-field to zero. The effects of a 
massive decay product have been investigated using the Hartree 
approximation in \cite{TKII}. For type~I 
models we will neglect the expansion of the Universe\cite{expanding}. 
For type~II
models our calculations are directly applicable to the expanding
Universe, as will be explained in section \ref{sec:massless}.

With the above approximations and {\it neglecting $\lambda_\chi$} the 
equations of motion for the modes of the $\chi$-field are 
\begin{equation} 
\frac{d^2\chi_k}{dt^2}+(k^2+g\phi^2(t))\chi_k=0\,. 
\end{equation} 
For type I models, where the oscillations of the $\phi$-field  EV are 
sinusoidal, this can be written as the Mathieu equation 
\begin{equation} 
\frac{d^2\chi_k}{dz^2}+(A(k)-2q\cos(2 z))\chi_k=0\,,\label{eq:mathieu} 
\end{equation} 
where $z=\omega_\phi t$, $A(k)=k^2/\omega_\phi^2+2q$, 
$q=g\Phi^2/4\omega_\phi^2$, $\omega_\phi=m$ is the frequency of the 
oscillations, and $\Phi$ is the slowly varying amplitude of the 
inflaton EV. For type II models the oscillations of $\phi$ are given 
by elliptic functions, but for the illustrative purpose at hand it is 
sufficient to replace the periodic EV by a sinusoid of the same 
frequency. In that case the $\chi$ modes again satisfy 
(\ref{eq:mathieu}), except that now 
$\omega_\phi=c\sqrt\lambda_\phi\Phi$, where $c\approx 0.85$.  The 
important point is that (\ref{eq:mathieu}) has exponentially growing 
solutions of the form $\chi_k \propto \exp(\mu^{(n)}_kz)$ within a set 
of resonance bands labeled by the integer index n. This growth of the 
modes corresponds to exponentially growing occupation numbers $n_k 
\propto \exp(2\mu^{(n)}_k\omega_\phi t)$ and may be interpreted as 
particle creation. Parenthetically we remark that for type II models
the inflaton may also decay into its own fluctuations. Writing
$\phi=\langle\phi\rangle+\delta\phi$ and linearizing one finds that the 
$\delta\phi_k$ also (approximately) obey (\ref{eq:mathieu}), with $A\approx
k^2/\omega_\phi^2+2.08$ and $q\approx 1.04$. For $g \gg \lambda_\phi$
the decay into $\chi$-fluctuations dominates, however, because
the relevant values of $\mu^{(n)}_k$ are a few times larger.
  
The literature distinguishes between narrow resonance, defined by $q 
\ll 1$, and broad resonance, for which $q \gtrsim 1$. While the first
case can be analyzed analytically 
\cite{ShtaTrashBran}, it is more difficult
to obtain quantitative results for the broad resonance case
\cite{KLS}. It is
precisely this regime, however, which is most interesting: 
models generally have $q \gg 1$ at the end of inflation, and it is 
broad resonance that leads to explosive particle creation.  To get 
a feel for the values of the parameters involved, note that 
typically $\Phi_0\sim M_{\rm P}$ at the end of inflation, where 
$M_{\rm P}=1/\sqrt{8\pi G}=2.4\times 10^{18}$~GeV
is the reduced Planck mass. 
COBE data \cite{CampbellDavidsonOlive} restrict
$m \lesssim 10^{-6}M_{\rm P}$ for type I models and
$\lambda_\phi \lesssim 10^{-12}$ for type II models. 
To prevent radiative corrections from
generating an inflaton self-coupling in conflict with COBE, one
generally needs  $g^2 \lesssim \lambda_\phi|_{max}$, i.e.\ $g \lesssim 
10^{-6}$. Finally, if $\chi$ is to represent a ``standard'' field then 
there is no reason for its self-interaction (as well as its coupling 
to other fields) to be tiny. Thus one might expect $\lambda_\chi \sim 
10^{-2} - 1$. The upshot of all this is that there is a ``natural'' 
hierarchy for the couplings in the model, 
\begin{equation} 
\lambda_\chi \gg g \gg \lambda_\phi\, ,\label{eq:couplings} 
\end{equation} 
and that q is generally large at the end of inflation. 
We believe that the role of $\lambda_\chi$ is of particular 
importance, especially since the final state self-coupling has been 
ignored in much of the literature so far. The exception is reference 
\cite{AllahverdiCampbell}, where it is was found that
the self-interactions can
have important effects in the case of narrow resonance. Below we will
see that the same holds true in the physically important broad
resonance regime.

The difficulty in analyzing the broad resonance case stems from the 
fact that all parameters in the Mathieu equation for the modes vary 
quite rapidly. For q very large particle production takes place during 
a tiny fraction of the period as the EV passes through zero. The 
amplitude $\Phi$ of the oscillations decreases as energy is 
transferred from the zero mode to the fluctuations. The produced 
$\chi$ particles generate a contribution to the mass of the inflaton 
of the form $g\langle\chi^2\rangle$, which changes $\omega_\phi$.  The 
$\chi$-field self-interaction, which we ignored in deriving the 
Mathieu equation for the fluctuations, produces an effective $\chi$ 
mass of the form $m^2_{\chi,\rm eff} \approx 
3\lambda_\chi\langle\chi^2\rangle$. This adds to $A(k)$ the time 
dependent term $m^2_{\chi,\rm eff}/\omega_\phi^2$. In addition to these 
{\it backreaction} effects there is scattering: the resonance produces 
particles in narrow momentum bins with large occupation numbers, and 
one certainly expects these to scatter and spread out rapidly. This
effect turns out to be very important, and it has not previously been
studied in two field models. 

\section{The method}

In order to take all of the above into account quantitatively we 
discretized the exact classical equations of motion and solved them on 
a three dimensional lattice. The classical equations of motion are a 
good approximation to the dynamics provided the mode amplitudes are 
large in the sense that the commutator of the canonical variables can 
be replaced by the Poisson brackets, which reduces to
$\phi_k\dot\phi_k \gg 1$ \cite{TK,FourierTransform}. 
Although our
initial state, which will be described below, does
not satisfy this condition, it is satisfied
very soon after the resonant decay begins.  
It is useful for the numerics to work in terms 
of dimensionless variables. We use
$\tilde{x}_\mu=\sqrt{g}\Phi_0 x_\mu$, 
$\tilde{m}=m/\sqrt{g}\Phi_0$, $\tilde{\lambda}_\phi= 
\lambda_\phi/g$, $\tilde{\lambda}_\chi=\lambda_\chi/g$, $\tilde{\phi} = 
\phi/\Phi_0$, and $\tilde{\chi} = \chi/\Phi_0$. In terms of these 
variables the initial inflaton EV amplitude $\Phi_0$ as well as $g$ 
completely disappear from the equations of motion. While $\Phi_0$ 
simply sets the scale in the problem and becomes irrelevant for the 
dynamics, the coupling $g$ is still important: it appears in the 
equations for the random initial fluctuations and regulates their 
amplitude compared to the EV.  The fields are evolved using an 
explicit algorithm which is second order accurate in time and fourth 
order accurate in space. We varied both the number of grid points and 
the lattice spacing to check that we are near the continuum limit. The 
data presented in this paper were obtained on $128^3$ 
lattices. Regarding numerical accuracy we note that the total energy 
of the two field system was conserved to better than $1\%$ in all of 
our runs. 
 
The initial conditions for the numerical simulation were chosen
according to the following reasoning: the initial amplitude of the
inflaton field is naturally of order the reduced Planck mass,
$\Phi_0\sim M_{\rm P} \equiv 1/\sqrt{8\pi G}=2.4\times
10^{18}$GeV\cite{endslowroll}. Typical values of the resonant momenta
are given by $A-2q\sim \sqrt{q}$, which leads to
\begin{equation}
k_{\rm res}\sim
\left(\frac{\omega_\phi \sqrt{g}\Phi_0}{2}\right )^{1\over 2}\simeq
\left( \frac{\sqrt{g}\Phi_0}{\omega_\phi} \right)^{1\over 2}
\frac{2 M_{\rm P}}{\Phi_0} H\simeq q^{1\over 4}
\frac{2 M_{\rm P}}{\Phi_0} H\,.
\end{equation}
For $q\gg 1$ this implies that $k_{\rm res}/ H\gg 1$ at the
end of inflation.  Since this ratio does not decrease at later times,
the resonant momenta are always on subhorizon scales.  These momenta
are not squeezed by inflation, and to a good approximation they are
zero point quantum fluctuations. The initial conditions may then be
approximated by a set of harmonic oscillators in the ground state.  We
define the phase space in terms of mode amplitudes
$\{\chi_k,\dot\chi_k\}$ for which $\langle \chi_k \rangle =\langle
\dot\chi_k\rangle=0$. These are defined as the minimum uncertainty
states with $\langle |\chi_k|^2 \rangle =1/(2\omega_k)$ and $\langle
|\dot\chi_k|^2\rangle=\omega_k/2$, where $\omega_k$ is the appropriate
mode frequency.  In the semiclassical approximation these states are
populated according to the phase space probability distribution.  In
our code we randomly generate gaussian distributed states with the
width given by $(2\omega_k)^{-1/2}$ for $\chi_k$ and
$(\omega_k/2)^{1/2}$ for $\dot\chi_k$, and similarly for the $\phi$
modes.  We point out that the evolution does not depend on details of
the initial conditions, as long as the initial field configurations
have the correct amplitudes \cite{Squeeze}.

\section{Massive inflaton (Type I models)}
\label{sec:massive}

\subsection{Models with $\lambda_\chi=0$}

Let us begin by discussing the simplest possible situation, namely a
type I model with $\lambda_\chi=0$. For the run presented here the
parameters were chosen as follows: $g=10^{-8}$,
$m^2=10^{-12}\Phi_0^2$, and lattice spacing $\Delta x=5\pi/4$. The
initial amplitude of the $\phi$-field EV, $\Phi_0$, sets the scale and
does not effect the dynamics in any way, as discussed above. Note that
for these values the q parameter in the Mathieu equation is $q=2500$,
putting us in the broad resonance regime.
 
Figure~1 shows the EV of the inflaton field $\phi$, which starts
decaying significantly around $t= 5000/\sqrt{g}\Phi_0=0.5\times
10^{8}\Phi_0^{-1}$. 
\begin{figure}[htb]
\epsfxsize=5in
\centerline{\epsfbox{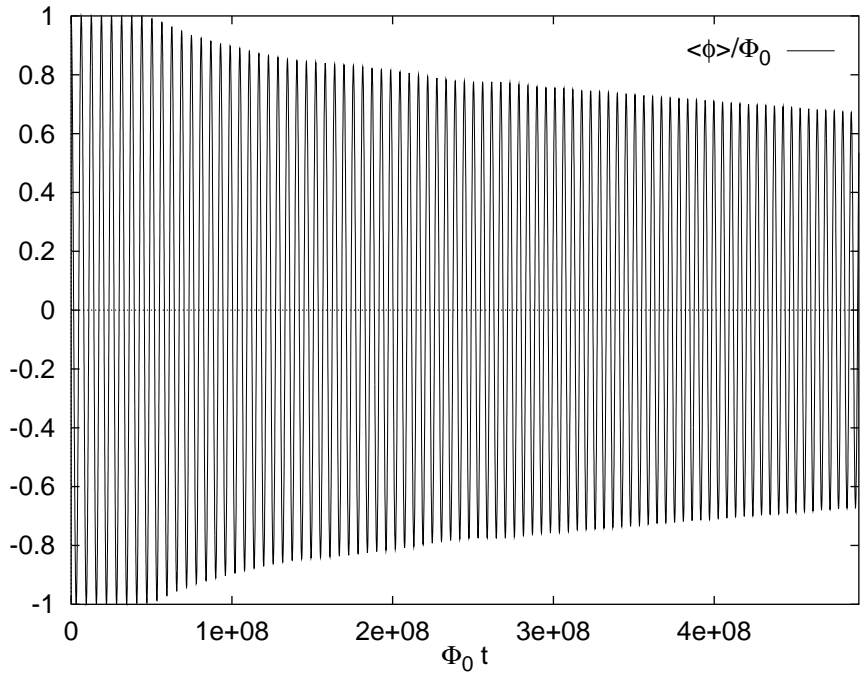}} 
\caption[The EV of the inflaton as a function of time]
{The EV of the inflaton as a function of time
($m=10^{-6}\Phi_0$, $\lambda_\chi=\lambda_\phi=0$, $g=10^{-8}$.)}
\end{figure}
Figure~2 shows the corresponding variances
$\langle(\delta\phi)^2\rangle =
\langle\phi^2\rangle-\langle\phi\rangle^2$ and $\langle\chi^2\rangle$
of the fields.
\begin{figure}[htb]
\epsfxsize=5in
\centerline{\epsfbox{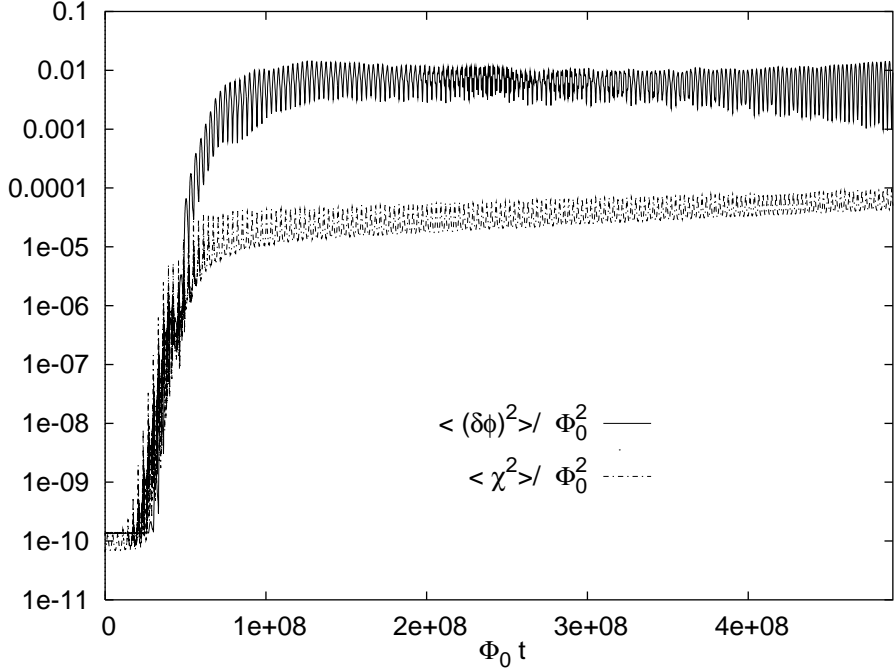}} 
\caption[The variances of the fields as a function of time]
{The variances of the fields as a function of time
($m=10^{-6}\Phi_0$, $\lambda_\chi=\lambda_\phi=0$, $g=10^{-8}$).}
\end{figure}
The variances increase roughly exponentially at
first. This trend stops when $\delta m^2_\chi\equiv g\langle (\delta
\phi) ^2\rangle$ becomes of order the resonant momentum squared. The
reason is that the $A(k)$ term in the Mathieu equation gets a
contribution of the form $g\langle (\delta \phi)
^2\rangle/\omega_\phi^2$. When this term becomes of order $k_{\rm
res}^2/\omega_\phi^2$ \cite{k-res}, the resonance band closest to
$A=2q$, which has the largest value of $\mu_k$, becomes inactive. In
the next band $\mu_k$ is down by a factor 3 \cite{mu-decay} so the
decay is slowed down dramatically.  The rapid production of inflaton
fluctuations, which kills the dominant resonance, may be surprising
since there is no $\phi$ resonance in this model.  These fluctuations
are produced by scatterings of resonant $\chi$-fluctuations off the
inflaton EV, which are fast, since both the $\phi$ zero mode and
$\chi$ resonant mode amplitudes are large.  It is interesting to note
that if one analyzes this model using a Hartree type approximation
which neglects scattering but includes backreaction effects, then the
condition for the end of explosive particle production is that
$q(t)\equiv g\Phi(t)^2/4(\omega_\phi^2+g\langle \chi^2\rangle)$
becomes $\ll 1$. We emphasize that this is {\it not\/} the correct
criterion: our simulations show that scatterings terminate the
exponential stage long before $q(t)\lesssim 1$. For example, one sees
from figures~2 and 3 that $q(t)\sim 600$ even at $t=5\times
10^8\Phi_0^{-1}$.
\begin{figure}[htb]
\epsfxsize=5in
\centerline{\epsfbox{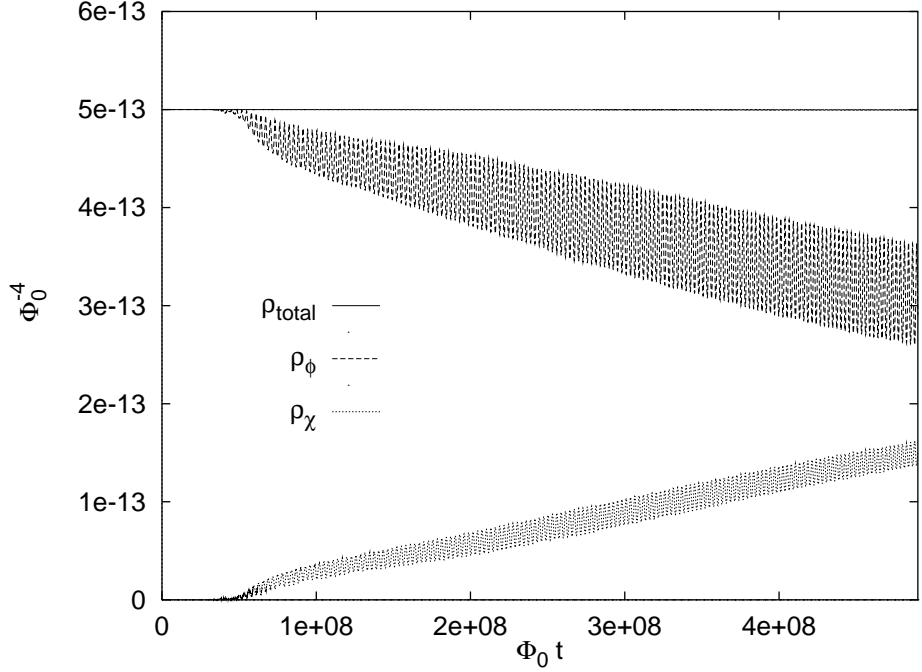}} 
\caption[The energy densities of the fields as a function of time]
{The energy densities of the fields as a function of time
($m=10^{-6}\Phi_0$, $\lambda_\chi=\lambda_\phi=0$, $g=10^{-8}$.)}
\end{figure}

The energy densities of the fields, as well as the total energy
density of the system, is shown in figure~3.  In the range $t\in
[0.8,5]\times 10^8\Phi_0^{-1}$ the $\chi-$field energy grows like
$t^\alpha$, with $\alpha \approx 0.95$ and a decay time scale of order
$\tau_{\rm decay}\sim 10^{9}\Phi_0^{-1}$.  Based just on the Mathieu
equation one would expect in this range exponential decay with time
scale
$\tau_{\rm decay}\sim 1/(2 \langle \mu_k\rangle \omega_\phi)
\sim 10^7\, \Phi_0^{-1}$. 
The reason is that as $\Phi$ decays, the resonances sweep
through the momentum space,
and as they move to higher momenta, $\mu_k$ diminishes, and new ones 
emerge in the infrared (IR). 
In estimating the above time scale, we have taken 
the typical value for $ \langle\mu_k \rangle\sim 0.05$.

To understand why the actual time scale is much longer it is useful
to look at the spectrum of fluctuations in k-space. To this end we
define ``occupation numbers'' for the $\chi-$field via
\begin{equation} 
n_k^\chi = 
\frac{\omega^\chi_k}{2}\langle\chi_{\vec k}\chi_{-\vec k}\rangle + 
\frac{1}{2\omega_k^\chi} 
\langle\dot{\chi}_{\vec k}\dot{\chi}_{-\vec k}\rangle\, ,
\end{equation} 
where $\omega^\chi_k=\sqrt{k^2+g\langle\phi^2\rangle}$, and the 
brackets denote averaging over directions. The 
$n_k^\phi$ are defined similarly, except that 
$\omega^\phi_k=\sqrt{k^2+m^2+g\langle\chi^2\rangle}$. Figure~4 
shows the occupation numbers of the $\chi$-field at various times. 
\begin{figure}[htb]
\epsfxsize=5in
\centerline{\epsfbox{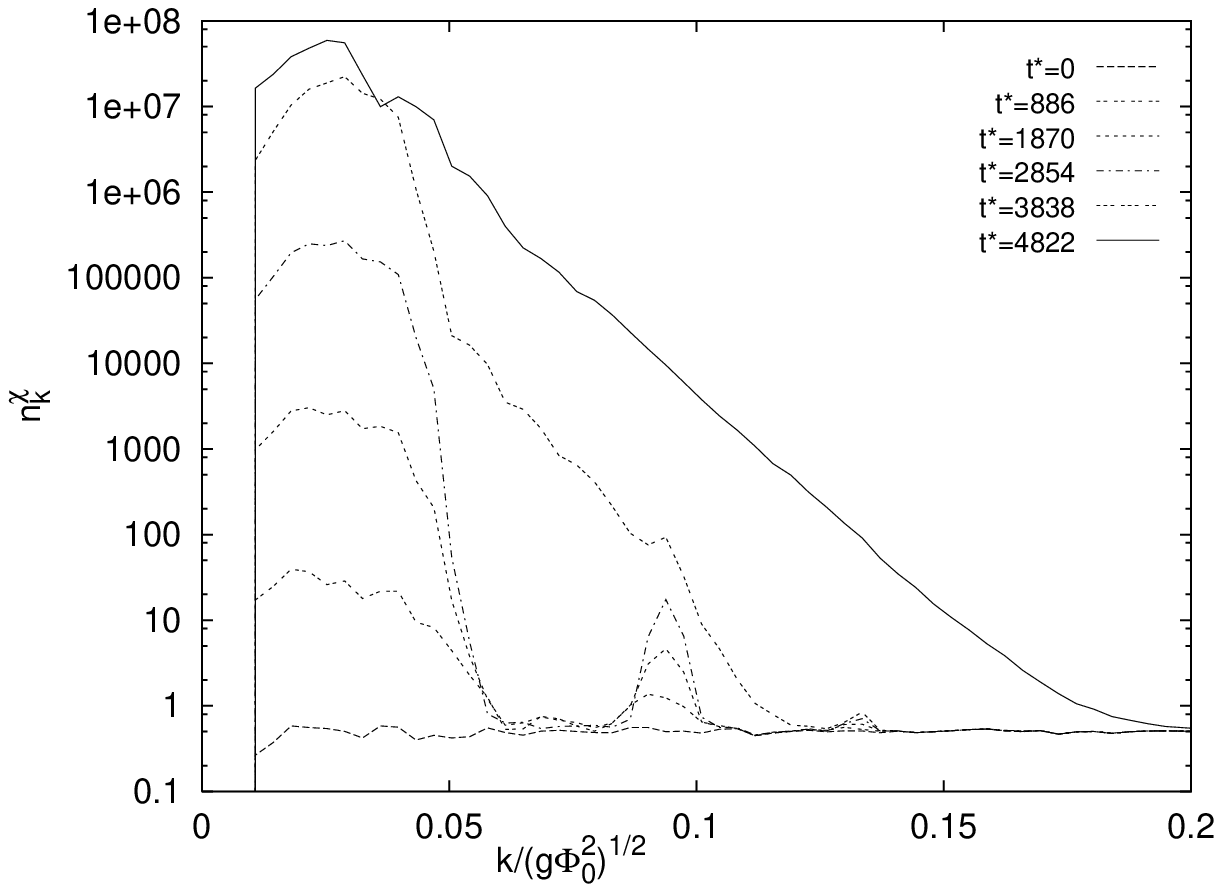}} 
\caption[$\chi$-field occupation numbers vs. momentum at early
times]{$\chi$-field occupation numbers vs. momentum at early
times. Three resonances are visible ($t^*=\protect\sqrt{g}\Phi_0 t $).}
\end{figure}
The
growth is exponential at first, with the lowest three resonance bands
clearly visible at $k\approx .025$, $k\approx .09$, and $k\approx
.13$, respectively. The values of $\mu_k^{(n)}$ obtained by assuming
$n_k^\chi \propto \exp 2\mu_k^{(n)}\omega_\phi t$ agree well with the 
predictions from the Mathieu equation. For the last time shown in 
figure~4, scatterings are already becoming important, and the 
resonance peaks get smeared out.

The subsequent evolution is shown in figure~5. 
\begin{figure}[htb]
\epsfxsize=5in
\centerline{\epsfbox{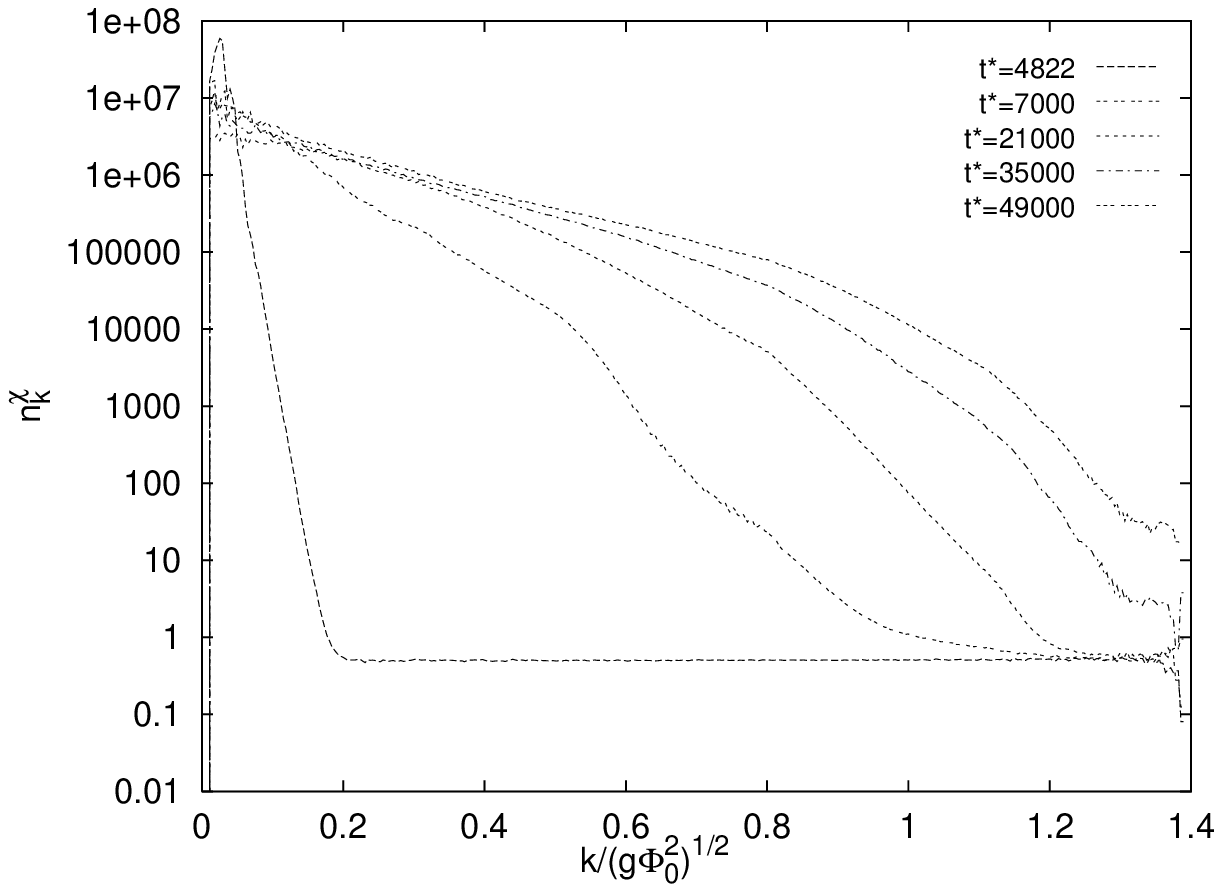}}
\caption[$\chi$-field occupation numbers vs. momentum at 
later times, when scatterings dominate]
{$\chi$-field occupation numbers vs. momentum at 
later times, when scatterings dominate ($t^*=\protect\sqrt{g}\Phi_0 t $).} 
\end{figure}
Scatterings are very 
fast after $t \approx 0.5\times 10^8\Phi_0^{-1}$, 
and the resonant peak structure is
quickly washed out. The occupation numbers grow to about
$n^\chi_k\sim 1/g$, and then remain approximately constant. 
There is a simple {\it feedback\/} mechanism which explains this 
behavior. Resonant particle production increases $n^\chi_{k_{\rm res}}$, 
which increases the scatterings off the zero mode, which  
increases infrared $\phi$ occupation numbers. This
increases $\delta m_\chi^2\equiv g\langle(\delta\phi)^2\rangle$, which 
decreases $\mu_k$ ({\it cf.\/} \cite{mu-decay}), slowing
down resonant particle production and giving scatterings
time to move particles toward higher momenta. This reduces
$\langle(\delta\phi)^2\rangle$, which increases $\mu_k$, increasing
resonant production and completing the feedback mechanism. 
Consequently $n_{k_{\rm res}}^\chi$ reaches an  
equilibrium value for which on average scatterings remove particles  
from the resonance as quickly as they are created, implying 
$\dot\rho_{\rm res}^\chi/\rho_{\rm res}^\chi= 
2\mu_k\omega_\phi-\Gamma_{\rm scatt}\sim 0$. 
The result is a slowly varying distribution with a characteristic 
shape, as seen at late times in figure 5. The slope 
$-k\, d\ln n^\chi_k/d k$ is small ($\lesssim 2$) in the infrared, 
grows as $k$ increases, and eventually becomes greater than 
$4$ at the scale $\Lambda$ at which 
most of the $\chi$ energy is concentrated.
Particles created by the resonance diffuse to this scale $\Lambda$ 
{\it via\/} scatterings and slowly move it to higher momentum. 
(Note that soon after scatterings become fast, $\Lambda\gg k_{\rm res}$.) 
Since the IR occupation numbers of the fields remain roughly constant  
during this process, the energy required is supplied by the zero mode.  
The number of $\phi$ zero mode ``particles'' eaten up in moving 
one $\chi$ particle from $k_{\rm res}$ to $\Lambda$ 
is about $N^\phi_0 \sim \Lambda/\omega_\phi\gg 1$, implying that  
scatterings are responsible for most of the EV decay.
The upshot is that energy is drained from the zero mode at a slowly 
varying rate given by  
$\dot{\rho_\phi}\sim N^\phi_0\rho^\chi_{\rm res}\, 2 \mu_k\omega_\phi$, where 
$\rho^\chi_{\rm res}$ is the energy density in the $\chi$ resonance. 
This implies a decay time scale 
\begin{equation}
\tau_{\rm decay}\sim \frac{ \rho_\phi}{\dot{\rho_\phi}}\sim
\frac{\rho_\phi}{\rho_{\rm res}^\chi\, 2\mu_k\omega_\phi}
\frac{\omega_\phi}{\Lambda}
\sim \frac{3\times 10^3}{g n_{k_{\rm res}}^\chi} 
\frac{\omega_\phi}{\Lambda} \frac{1}{\omega_\phi}\,.
\label{eq:decay time}
\end{equation}
To obtain the last expression in Eq.~(\ref{eq:decay time}) we used
\begin{equation}
\rho_{\rm res}^\chi \simeq \int_{\rm res}
\frac{d^3 k}{(2\pi)^3} n_k^\chi\omega_k^\chi \sim \frac{k_{\rm
res}^3}{2\pi^2}
w_{\rm res} n_{k_{\rm res}}^\chi\,,
\end{equation}
where the resonance width $w_{\rm res}$ is approximately given by 
\begin{equation}
w_{\rm
res} \simeq 0.8 k_{\rm res} \exp{[-(m_\chi^2+k^2)/(\Delta k)^2]}\, ,
\end{equation}
as well as $k_{\rm res}^2$ $\simeq$ $\sqrt{q}\omega_\phi^2=
\sqrt{g} m_\phi \Phi/4$, and $\mu_k\simeq 0.23
\exp{[-(m_\chi^2+k^2)/(\Delta k)^2]}$.  
Here $\Delta k\simeq (\sqrt{g} m_\phi \Phi/2)^{1/2}$ 
denotes the distance between the first two 
resonances. The numerical coefficient is obtained by noting that 
resonant energy production peaks around
$k\simeq k_{\rm res}$, and 
$m^2_\chi\equiv g\langle(\delta\phi)^2\rangle\sim k_{\rm res}^2$.
The last relation remains reasonably well satisfied throughout 
the scattering regime. This is not surprising, since from 
$2\mu_k\omega_\phi\sim \Gamma_{\rm scatt}$ one finds
$m_\chi^2\sim k_{\rm res}^2\ln (\Gamma_{\rm scatt}/2\mu_0\omega_\phi)$,
{\it i.e.\/} $m_\chi^2$ is only logarithmically dependent on slowly 
varying quantities (from above, $\mu_0\simeq 0.2/e$).  
The quantities entering Eq.~(\ref{eq:decay time}) 
vary slowly during the scattering dominated regime, and one
should take their typical values to get an order of 
magnitude estimate for the decay time. 
From figure 5 one reads off
$g n_{k_{\rm res}}^\chi\simeq 0.04$ and 
$\Lambda/\omega_\phi\sim 70$, which leads to 
$\tau_{\rm decay}\sim 10^9 \Phi_0^{-1}$.
This agrees with the decay time scale seen in figures 1 and 3.  
The time scale in (\ref{eq:decay time})
is naively parametrically larger by $\sim 1/[g\ln (1/g)]$
in comparison to the scale of parametric resonant 
decay. The reason why the decay time is not huge even for $g$ 
very small is that $g n^\chi_{k_{\rm res}}\sim 0.04$, {\it i.e.\/},
the occupation numbers are very large. 
This however is not true when there is a 
moderate self-coupling of the $\chi$ field, as will be discussed in
section~\ref{subsec:large lambda chi}.

From the discussion above it is clear that the inflaton 
decay consists of two distinct regimes: fast exponential
decay during which scatterings are irrelevant followed by 
a much slower decay governed by scattering processes. 
One can clearly separate these two regimes in figure 2:
the scattering regime sets in at the time when the variances 
become slowly varying, which occurs at 
$t\simeq 0.8\times 10^8\Phi_0^{-1}$. 
Clearly there is a brief transition stage between the two regimes, 
and since the exponential decay is a stimulated process, this is 
where most of the energy loss associated with the exponential regime
occurs.
The resonant occupation numbers stay 
roughly constant during this transition stage since the
scatterings are already fast, but the infrared and near ultraviolet
states are quickly filled.
One can estimate quite generally what fraction of the 
inflaton energy has decayed at the time when the slowly varying state
sets in,
as follows. We have run our code for several values of $g$ in the 
range $g=10^{-12}$ to $10^{-8}$ and find that
$g\langle \chi^2\rangle
\simeq g\int [d^3 k/(2\pi)^3 \omega^\chi_k ] n^\chi_k
\sim \omega_\phi^2/4$ when 
the scattering regime begins 
\cite{equipartition}.
This value is somewhat dependent on the exact initial position of the 
dominant resonance.
Since at this time the variance is dominated by modes 
with $k^2\sim \sqrt{g}\omega_\phi\Phi_0$,
we can estimate the occupation numbers of the $\chi $ field
around the resonant momentum
to be $g n^\chi_{k_{\rm res}}\sim 5 \omega_\phi/\sqrt{g}\Phi_0$. 
At this time the $\chi$ energy is dominated 
by approximately the same scale 
$( \sqrt{g}\omega_\phi\Phi_0)^{1/2}$, implying 
$\rho_\chi/(\rho_\phi)_0\sim \omega_\phi/\sqrt{g}\Phi_0
=(1/2){q}^{-1/2}$.
This means that for $q\gg 1$ only a small fraction 
of the inflaton energy decays during the exponential regime
\cite{m-expanding-case}.

After the scatterings become important, the distribution broadens
significantly, but the occupation number at the resonant 
scale remains to a good approximation constant. 
With our above estimate for $gn^\chi_{k_{\rm res}}$ we can thus 
give a parametric expression for the decay time in the scattering 
regime based on Eq.~(\ref{eq:decay time}).
Our numerical results indicate that occupation numbers 
drop approximately exponentially between $k_{\rm res}$ and 
$\Lambda$ ({\rm cf.\/} figure 5).  We take
$n_{k}^\chi=n_{k_{\rm res}}^\chi\exp{[-(k-k_{\rm res})/\kappa]}$,
where $1/\kappa$ is the slope of $\ln n_k^\chi$. As can 
be seen in figure 5, this value decreases slowly as the field decays.
The relevant value of $\kappa$ 
can be computed by noting that 
when the EV decays significantly, 
about half of the original energy density
$(\rho_\phi)_0=\omega_\phi^2\Phi_0^2/2$ 
is in the $\chi$ fluctuations. This
gives $\kappa^2\simeq q^{1/4}\sqrt{g}\Phi_0\omega_\phi/2$.
The scale $\Lambda$ can be evaluated as 
$\Lambda=\langle k \rangle \simeq 4\kappa$.
Combining these results with Eq.~(\ref{eq:decay time})
we arrive at the following parametric estimate
for the decay time
(recall that this is sensible only for $q\gg 1$, since otherwise most 
of the field decays during the exponential regime):

\begin{equation}
\tau_{\rm decay} \sim 
3\times 10^2 q^{\frac{1}{8}}\frac{1}{\omega_\phi}
\,,\qquad 
q\gg 1
\label{eq:decay time II}
\end{equation}
The main point of this equation is that the decay time is 
very weakly dependent on $q$.
(This conclusion is independent of our assumption that
$n_k^\chi$ falls off exponentially in the 
relevant momentum range, even though the exact power of $q$ is not.)
Note that surprisingly, for 
fixed $\omega_\phi$ and $\Phi_0$, the decay time 
actually slowly increases as $g$ increases. 
That $\tau_{\rm decay}$ is nearly independent
of $q$ in the scattering regime agrees well 
with our numerical simulations for $2.5\times 10^3>q>25$. 
We have not verified this dependence for much larger values of
$q$ because such simulations require enormous
computing times.
We showed above that the fraction of energy that decays in 
the exponential regime is $\sim 1/\sqrt{q}\propto 1/\sqrt{g}$,
implying that paradoxically the field decays faster for
smaller values of $q\propto g$. 
Our numerical calculations indeed indicate that the fastest 
inflaton decay occurs for $q\sim 1$, rather than for $q\gg 1$. 
One should note, however, that in an expanding universe
parametric resonance is ineffective for $q\sim 1$
in type I models since the effective $q_t$ becomes much less that unity
before the resonant occupation numbers grow large
\cite{TKII}.

\subsection{Models with $\lambda_\chi\gg g$}
\label{subsec:large lambda chi}

We now investigate what happens when $\lambda_\chi \gg g$. As discussed
in the introduction, this occurs naturally in many models.
For the run to be presented we used $m^2=10^{-12}\Phi_0^2$, 
$g=10^{-8}$, $\lambda_\phi=0$, $\lambda_\chi=0.01$, and $\Delta x=5\pi/2$. 
The evolution in this case is very different from the $\lambda_\chi=0$
case (case I$(a)$) considered
above. As can be seen from figure 6, the resonance starts growing 
just as in case  I$(a)$. 
\begin{figure}[htb]
\epsfxsize=5in
\centerline{\epsfbox{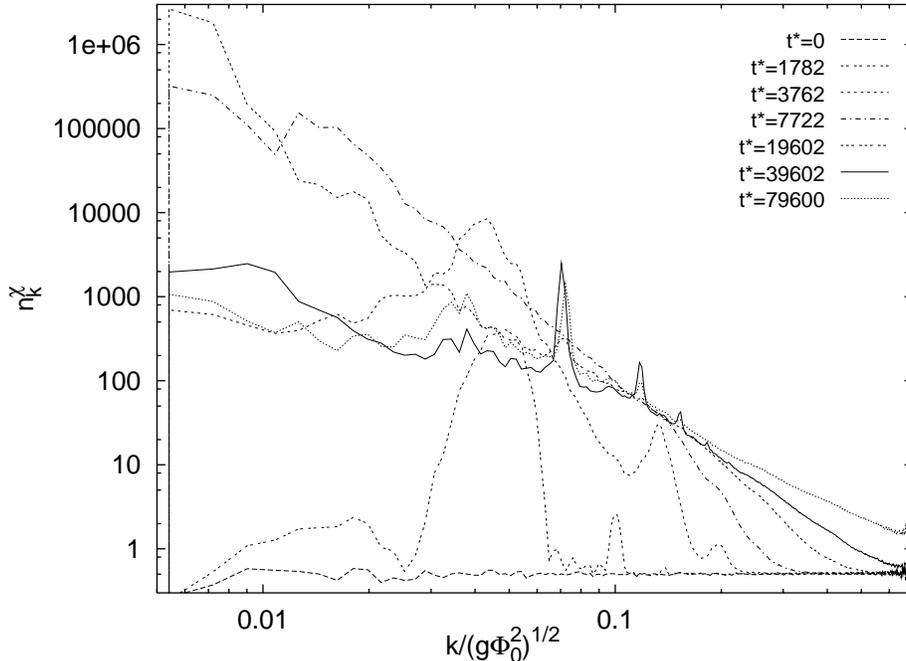}} 
\caption[$\chi$-field occupation numbers at various times for a model
with moderate self-coupling]
{$\chi$-field occupation numbers at various times for a model
with moderate self-coupling $\lambda_\chi=10^{-2}$ ($m=10^{-6}\Phi_0$,
$\lambda_\phi=0$, $g=10^{-8}$ and $t^*=\protect\sqrt{g}\Phi_0 t $).}
\end{figure}
However, when the occupation numbers reach 
about $10^2/\lambda_\chi$, scatterings become fast and 
populate the infrared states. This gives rise to 
the backreaction $\delta m^2_\chi=3\lambda_\chi \langle \chi^2\rangle$,
which shifts the resonances toward the infrared. As the first resonance 
sweeps through the infrared, it further populates the states. 
Once the resonance reaches $k\sim 0$, it is no longer very effective
so that scatterings remove particles from the infrared
faster than they are supplied by the resonance. This leads to 
depletion of the infrared states, which decreases the backreaction
a little, increasing the resonance's effectiveness.
This feedback mechanism leads to a state with slowly varying
occupation numbers, just as in case I(a). 
As a consequence, energy is removed from the zero mode at a 
slowly varying rate, starting at about $t=2.5\times 10^8\Phi_0^{-1}$. 
In contrast to case I(a) the $\phi$ field plays no part in keeping the
occupation numbers constant: it is the much faster $\chi - \chi$
scatterings which remove particles from the resonant momenta as they
are created \cite{I(b)}.  
Consequently the EV does not get depleted by scatterings, and the
rate of energy density loss of the $\phi$-field is simply
$\dot \rho_\phi\simeq -\rho_{\rm res}^\chi 2\mu_k \omega_\phi$,
which leads to an estimate of the decay time
$\tau_{\rm decay}=-\rho_\phi/\dot\rho_\phi\simeq
10 \omega _\phi\Phi_0^2/
(2\mu_k k_{\rm res}^3 w_{\rm res} n_{k_{\rm res}}^\chi)$.
The parameters can be deduced from our simulation. The relevant 
resonance in this case is the second one (recall that the first one 
has been shifted to the infrared) which can be seen at 
$k \approx 0.07 \sqrt{g}\Phi_0$ at late times in figure~6. We estimate   
$k_{\rm res}^2\simeq 0.5 \sqrt{g}\omega_\phi \Phi_0$, 
$\mu_{k_{\rm res}}\simeq 0.03$, the resonance 
width $w_{\rm res}\simeq 0.05 k_{\rm res}$, and 
$n_{k_{\rm res}}^\chi\simeq 15/\lambda_\chi$, so that 
$\tau_{\rm decay}
\sim 10^{4}/(gn_{k_{\rm res}}^\chi \omega_\phi)
\sim 10^{15}\Phi_0^{-1}$. 
Note that this time scale is parametrically larger than 
the resonance decay time by  
$\lambda_\chi/g \ln(1/g)$, which is about $10^5$ in our case. 
We point out that $\langle \chi^2\rangle$ reaches a maximum value
of about $6\times 10^{-9}\Phi_0^2$ at 
$t\sim 2\times 10^8\Phi_0^{-1}$, and by 
$t=8\times 10^8\Phi_0^{-1}$ less than 
$0.0005\%$ of the energy is in the $\chi$ field
\cite{chi-energy}.
The rate of energy density increase is about 
$\delta \rho_\chi/\delta t \sim 4\times 10^{-28}\Phi_0^5$.
A linear extrapolation gives for the decay time 
$\tau_{\rm decay}\sim \rho_\phi /(\delta \rho_\chi/\delta t)
\sim 10^{16}\Phi_0^{-1}$, in rough agreement
with our simple model.  We emphasize that we do not claim that 
at very late times the simple feedback mechanism gives a correct 
picture. The main point of this simulation was to  
establish that a moderate self-coupling of the $\chi$ field
slows down the EV decay by many orders of magnitude. 
This conclusion remains valid when $\chi$ couples moderately to 
other fields. The only difference is that $\chi$ now 
decays into these fields, which in turn induces 
backreaction on $\chi$. While the details are model dependent,
the basic mechanism remains valid, and parametric resonance is 
rendered ineffective.

\section{Massless inflaton (Type II models)}
\label{sec:massless}

Let us now turn to type~II models. These were investigated using the 
same techniques as above, so rather than present the results in detail 
we will simply state our conclusions. Before doing so it is worth 
pointing out that for these models the expanding universe equations of 
motion are conformally equivalent to those in Minkowski space. By this 
we mean that in terms of the variables $\tau=\int dt/a(t)$, 
$\bar{\phi}=\phi a(\tau)/a(0)$, and $\bar{\chi}=\chi a(\tau)/a(0)$, 
where $a(\tau)$ is the scale factor, the Friedmann-Robertson-Walker
(FRW) equations of motion are
of exactly the same form as those in ordinary static
spacetime \cite{static vs expanding}. Hence, with the simple 
replacements above, our
numerical calculations for type~II models apply directly to an
expanding universe.

The main results of our study are as follows.

\subsection{Models with $\lambda_\chi=0$}
\label{sec:massless coupless}
 If $\lambda_\chi=0$,
the inflaton decays via parametric resonance in much the same way as
it did in case I(a). That is, after a brief period of exponential
decay a slowly varying state sets in during which the decay is
dominated by scatterings. 
For example, figure 7 shows the variances of the fields
for $m^2=0$, $\lambda_\phi=10^{-12}$, and $g=10^{-8}$.
\begin{figure}[htb]
\epsfxsize=5in
\centerline{\epsfbox{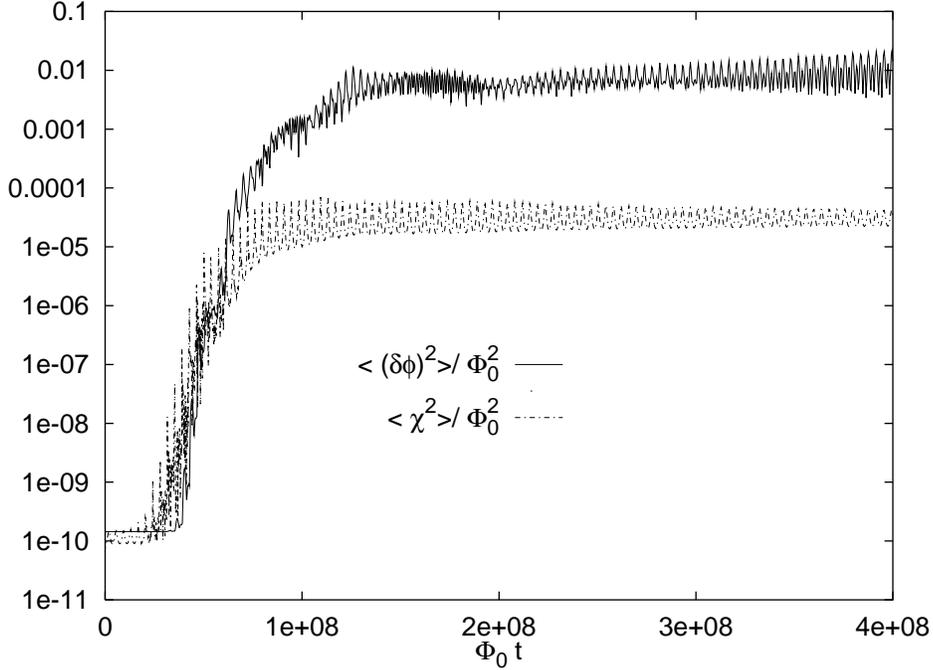}} 
\caption[The variances of the fields as a function of time
for a quartic inflaton potential]
{The variances of the fields as a function of time
for a quartic inflaton potential ($m=0$,
$\lambda_\phi=10^{-12}$, $g=10^{-8}$).}
\end{figure}
Since in this case 
$\omega_\phi=0.85\sqrt{\lambda_\phi}\Phi_0$, $q\approx 3500$.
The figure indicates that, as in case I(a), the 
exponential regime ends at $t\sim 0.8\times 10^8\Phi_0^{-1}$. 
The maximum values of the variances
are again to a good approximation 
given by $g\langle \chi^2\rangle \sim \omega_\phi^2$
and $g\langle \phi^2\rangle \sim \omega_\phi\sqrt{g}\Phi_0$.
The decay time can still be estimated from
(\ref{eq:decay time}), with the replacement $m_\phi \rightarrow
\omega_\phi\simeq 0.85\sqrt{\lambda}\Phi_0$. 
The slowly varying value of $gn^\chi_{\rm{k_{res}}}$ is again about 
$5\omega_\phi/\sqrt{g}\Phi_0$, so type I and II
models behave very much alike for equal values of the parameters.
One difference is that for type II models
the inflaton energy decays somewhat faster, since 
$\mu_k$ is larger and in addition
there are scatterings via the $\lambda_\phi\phi^4$ term.
For the $\lambda_\chi=0$ runs described in this work,  
the energy decays about $50\%$ faster for the quartic potential. 
The value of $\mu_k$ is about $25\%$ larger for that case.

As discussed at the beginning of this section, the lattice results for
type II models can be mapped onto the expanding universe by the
appropriate rescaling. 
To see how this works in practice
consider the problem of obtaining 
the maximum value of $\langle \chi^2\rangle$ reached
during preheating in the expanding universe. This value is interesting
because it quantifies the effective temperature of non-thermal phase
transitions \cite{nonthermalPT}. Since for type II models
the universe is radiation dominated we can write 
$a(\tau)/a(0)=1+H_0a(0)\tau$,
where $\tau$ is the conformal time and 
$H_0=\sqrt\lambda_\phi\Phi_0^2/\sqrt{12}M_{\rm P} $ is the 
Hubble constant at the end of inflation. As discussed at the beginning 
of section~\ref{sec:massless}, 
in a Friedmann-Robertson-Walker universe our lattice fields are 
rescaled by a factor $a(\tau)/a(0)$ and our lattice time is $\tau$. 
Hence $\langle \chi_{\rm\; FRW}^2\rangle = 
(a(0)/a(\tau))^2\langle \chi^2\rangle$. Consider, for example, 
the variances shown in
figure~7. Since $\langle \chi^2\rangle$ rises rapidly
and then becomes slowly varying, 
$\langle \chi_{\rm\; FRW}^2\rangle_{\rm max}$
occurs at the moment when the fast growth terminates. This occurs
at $\tau\approx 8000/\sqrt g\Phi_0=8\times10^7 \Phi_0^{-1}$ when $\langle
\chi^2\rangle \approx 6\times 10^{-5}\Phi_0^2$. Hence we
obtain $\langle \chi_{\rm\; FRW}^2\rangle_{\rm max} \approx 6\times
10^{-5}\Phi_0^2/(1+2\times10^7 
\sqrt\lambda_\phi\Phi_0/M_{\rm P})^2\approx
10^{-7} M_{\rm P}^2$, 
where we have used
$a(0)=1$ in our units. We can obtain general formulae for the maximum
variances of the fields when $q\gg 1$ by recalling that the slowly varying
scattering regime sets in when  $g\langle
(\delta\phi)^2\rangle \sim k_{\rm res}^2\sim
\sqrt g\omega_\phi\Phi_0/2$ and $g\langle
\chi^2\rangle \sim \omega_\phi^2/4$, where
$\omega_\phi\approx 0.85 \sqrt\lambda_\phi\Phi_0$. This occurs when
$n^{\chi}_{k_{\rm res}}\sim (g\sqrt q)^{-1}$, {\it i.e.\/}
at the time $\tau \sim \ln(1/g\sqrt q)/2\mu_k\omega_\phi$. 
Combining this with our
equation for $a(\tau)$ we obtain 
\begin{mathletters}
\begin{eqnarray}
\langle
(\delta\phi_{\rm \;FRW})^2\rangle_{max} 
& \sim & 0.2\frac{1}{q^{1/2}\ln^2(\lambda_\phi^{-1} q^{-3/2})}M_{\rm P}^2
\\
& \sim & \left (\frac{\lambda_\phi}{g}\right )^{1\over 2}
\frac{1}{\ln^2(\lambda_\phi/g^3)}M_{\rm P}^2
\end{eqnarray}
\end{mathletters}
and
\begin{mathletters}
\begin{eqnarray}
\langle \chi_{\rm\; FRW}^2\rangle_{max} 
& \sim & 0.06\frac{1}{q\ln^2(\lambda_\phi^{-1} q^{-3/2})}M_{\rm P}^2
\\
& \sim & \left (\frac{\lambda_\phi}{g}\right )
\frac{1}{\ln^2(\lambda_\phi/g^3)}M_{\rm P}^2\;,
\end{eqnarray}
\end{mathletters}where $M_{\rm P}\approx 2.4\times 10^{18}$GeV.
These results can be considered upper limits for the variances since
they were obtained for a massless $\chi$ field and a mass term in the
equations of motion only suppresses the resonance
\cite{TKII}. Note that 
$\langle\chi_{\rm\; FRW}^2\rangle_{max} \propto q^{-1}$ 
(up to logarithms), while Hartree
type approximations which neglect scattering predict a $q^{-1/2}$
dependence \cite{KLS}. That
the Hartree approximation may not give the correct $q$ dependence
of $\langle\chi^2\rangle$ was previously mentioned in 
\cite{TKII}.

\subsection{Models with $\lambda_\chi\gg g$}

If $\lambda_\chi \gg g$, the situation is drastically changed. Just as
for type I models the decay into the $\chi$-field is extremely
slow. But for type~II models the inflaton can decay into its own
fluctuations via parametric resonance, as discussed below 
Eq.~(\ref{eq:mathieu}). This process, which is much slower than decay into
$\chi$ fluctuations for $\lambda_\chi=0$, becomes dominant
for $\lambda_\chi\gg g$. In fact the inflaton decays into its own
fluctuations as if it were not coupled to the $\chi$-field at all. We have
verified this by running the one field case and comparing the results
to the two field simulation with the same parameters in the $\phi$
sector. For example, with $m=0$, $\lambda_\phi=10^{-12}$,
$\lambda_\chi=10^{-2}$ and
$g=10^{-8}$ we find that $73\%$ of the energy has decayed by
$t=8\times10^{8}\Phi_0^{-1}$ and $\langle (\delta\phi)^2\rangle_{\rm max}
\approx 8\times 10^{-2}\Phi_0^2$, which agrees very well with 
the corresponding one field run. We also note that
in the two field run $\langle \chi^2\rangle$ grows exponentially
at first but then reaches a maximum value of about 
$3\times 10^{-9}\Phi_0^2$ at $t\sim 0.4\times 10^8\Phi_0^{-1}$.
The details of the one field case are discussed in
\cite{TK}.
One may wonder why the perturbative energy flow via scatterings
from $\phi$ to $\chi$ fluctuations is ineffective even after 
the infrared $\phi$ occupation numbers reach a huge value
of order $\lambda_\phi^{-1}\gg g^{-1}$. The reason is that 
due to the backreaction $g\langle \delta\phi^2\rangle\;$, 
$\chi$ fluctuations become massive and 
$\phi\phi\rightarrow\chi\chi$ scatterings are kinamatically forbidden.
(Indeed, taking the above value for 
$\langle (\delta\phi)^2\rangle_{\rm max}$,
one finds $m_\chi/\omega_\phi\sim 30$.) 

We point out that it is very difficult to simulate type~II models with
large q on the lattice. The reason is that the typical resonant
momenta for the two fields differ by a factor $k_{\rm res}^\phi/k_{\rm
res}^\chi \sim q^{-1/4}$, making it difficult to run simulations that
capture both resonances with sufficient infrared and ultraviolet
resolution. Our numerical results on $128^3$ lattices were hence
obtained in a two step process: we first did preliminary runs,
choosing $\Delta x$ optimally for each resonance in turn. This allowed
us to determine which one was dominant, namely the $\chi$ resonance
for $\lambda_\chi=0$ and the $\phi$ resonance for $\lambda_\chi\gg g$.
We then did extended runs choosing $\Delta x$ just large enough to
capture the physics of the dominant resonance, giving us the best
possible ultraviolet range for the given situation.

\section{Conclusion}

We have numerically investigated the decay of the 
inflaton coupled to a massless scalar field. Our main results 
for both quadratic (type I) and quartic (type II) inflaton potentials
are as follows: 

\noindent 1. If the self-coupling of the decay product is small 
($\lambda_\chi\ll g$) we find two distinct stages of 
inflaton decay for $q \gtrsim 25$: an exponential regime in 
which the field decays via parametric resonance, followed by a
scattering dominated regime. The fraction of energy which decays
by the time the scattering regime begins
is of order $q^{-1/2}$. This means that 
for $q\gg 1$ scatterings are responsible for most of the inflaton 
decay, and we find that the decay time scale in the scattering
regime is significantly longer than the resonance decay time scale.

\noindent 2. If 
$\lambda_\chi\gg g$ (which is natural for many models) we find that 
parametric resonance cannot transfer energy to $\chi$.
The reason is that scatterings due to the self-interactions
limit the maximum occupation numbers of $\chi$, allowing only a tiny 
fraction of the inflaton energy to be transferred during
the resonant stage.  
In the subsequent scattering dominated regime the transfer of
energy is extremely slow, leading to huge decay times for type I 
models. For type II models the inflaton decays into its own
fluctuations, essentially as if it were not coupled to the 
second field at all. Although these results were obtained  
in a simple model with two scalar fields, 
the basic mechanism, and hence our conclusion, remains 
valid for a realistic model with many mutually interacting
fields.

In this work we have studied inflaton decay by integrating the
Minkowski space equations of motion.  It is an interesting question
how our conclusions might change in an expanding universe.  For type
II models the answer is simple: as explained in
section~\ref{sec:massless}, our analysis is directly applicable after
the appropriate rescaling. The mapping was made explicit in
section~\ref{sec:massless coupless}, where we obtained upper limits
for the variances of the fields in the expanding universe.

For type I models the situation is more complex. 
The time scale $H^{-1}$ defined by the the expansion rate 
$H=\sqrt{\rho_\phi/3}/M_{\rm P}\lesssim
\omega_\phi/2$
is much shorter than the scattering decay time scale (see
Eq.~(\ref{eq:decay time II})).  Consequently we cannot expect our
work for type I models to be a good approximation to the expanding
universe case.  However, we do expect our conclusion that the inflaton
cannot decay into a field with moderate self-interactions to remain
valid. First of all, parametric resonance in an expanding universe is
suppressed. Second, in the scattering regime the maximum occupation
number will remain of order $\lambda_\chi^{-1}$, so one expects that
the total energy deposited in the $\chi$ field fluctuations remains a
tiny fraction of the inflaton energy even in an expanding universe.

\acknowledgements This research was conducted using the resources of
the Cornell Theory Center, which receives major funding from the
National Science Foundation (NSF) and New York State, with additional
support from the Advanced Research Projects Agency (ARPA), the
National Center for Research Resources at the National Institutes of
Health (NIH), IBM Corporation, and other members of the center's
Corporate Partnership Program.  We would like to thank Robert
Brandenberger, Claudia Filippi, Brian Greene, Andrei Linde, and Guy
Moore for suggestions and helpful discussions. TP acknowledges
funding from the U.S. NSF.


\begin{thebibliography}{99}
\bibitem{endslowroll}
The end of inflation is signaled by a breakdown of the slow-roll
condition $\ddot\phi\ll 3H\dot\phi$. The relation $\ddot\phi\sim
3H\dot\phi$ then leads to $\Phi_0\sim M_{\rm P}$ for both $m\ll
\sqrt{\lambda_\phi}\Phi_0$ and $m\gg
\sqrt{\lambda_\phi}\Phi_0$.
\bibitem{Squeeze} Our initial conditions differ from those used in
\cite{TKII} in that our field amplitudes and
``velocities'' are not correlated.  The authors of \cite{TKII} obtain
correlated initial conditions by evolving the developed resonant
solution backward in time, while we simply take the initial state to
correspond to random zero point fluctuations.  We agree with the
authors that the details of the initial conditions do not affect the
evolution in any significant way, so long as they have the correct
amplitudes. We have verified this explicitly by reproducing the
results of \cite{TK} using our initial conditions.
\bibitem{k-res} The initial position of the dominant resonance 
depends on the exact value of $q$, and can be anywhere between
$k=0$ and $k=\Delta k_{\rm res}$, where 
$\Delta k_{\rm res}^2\approx \sqrt{q}\omega_\phi^2
=\sqrt{g}\Phi_0\omega_\phi/2$ is the typical distance between the 
first two resonances.
\bibitem{mu-decay}{
We find  that for the first few resonances above 
$A=2q$ the maximum $\mu_k$ decays approximately as 
$\mu_k\propto \exp {[-\alpha (k^2+m_{\chi\,,{\rm eff}}^2) 
/\Delta k_{\rm res}^2]}$, where 
$\Delta k_{\rm res}^2\simeq \sqrt{g}\omega_\phi\Phi_0/2$ 
is the distance between the resonances, and 
$\alpha \simeq 1.1$. The width of the resonances is also decreasing 
exponentially. 
}
\bibitem{equipartition} 
This observation can be understood as follows. At the beginning of the 
scattering regime the flow of energy between
the two fields is fairly efficient since the scatterings are already fast.
Thus one might expect that equipartition of energy 
between $\phi$ and $\chi$ fluctuations is reasonably well satisfied. 
(We typically find that at this stage the energy in $\phi$ fluctuations
is several times larger than that in $\chi$ fluctuations.
At later stages equipartition is well satisfied.)
From $\rho_{\delta\phi}\sim \rho_\chi$ 
one finds $n_\chi\sim n_\phi (\omega_\phi/\Lambda')^4$, where 
we have assumed that the energy in the $\phi$ and $\chi$ 
fluctuations is concentrated at the scales  $\omega_\phi$
and $\Lambda'\sim (\sqrt{g}\Phi_0\omega_\phi)^{1/2}$, respectively. 
Recall that the backreaction renders the dominant resonance inactive when
$g\langle (\delta\phi)^2\rangle\sim k_{\rm res}^2
\sim \sqrt{g}\Phi_0\omega_\phi/4$, from which one obtains 
$n_\phi\sim (2\pi^2/g) (k_{\rm res}/\omega_\phi)^2$.
Combining these estimates, we find
$n_\chi\sim (\pi^2/2g) \omega_\phi/\sqrt{g}\Phi_0$ and hence 
$g \langle \chi^2\rangle\sim \omega_\phi^2/4$.
\bibitem{m-expanding-case}
One may wonder how this changes in
an expanding universe. In terms of 
the variables $\tilde\phi=\phi (a/a_0)^{3/2}$ and 
$\tilde\chi=\chi (a/a_0)^{3/2}$, where
$(a/a_0)^{3/2}\simeq (3/2\sqrt{6})(\Phi_0/M_{\rm P})m_\phi t$,
the $\tilde\chi_k$ obey an approximate Mathieu equation,
with time dependent parameters $A_t$ and 
$q_t=(g\Phi_0^2/4m_\phi^2) (a_0/a)^3$. 
Since $H(t)/m_\phi=(\Phi_0/\sqrt{6}M_{\rm P})(a_0/a)^{3/2}\ll 1$ 
(except very early when it is $\lesssim 1/2$),
$q_t$ changes slowly and the adiabatic approximation should be 
reasonably accurate. The scattering regime begins at 
$t\sim \ln g^{-1}/2\langle \mu_k\rangle m_\phi$, which can be used to
compute $(a/a_0)^{3/2}\simeq 10^{-1}
(\Phi_0/M_{\rm P})^2(\ln g^{-1}/ \langle \mu_k\rangle) ^2$.
One may go through a similar resoning as in the Minkowski case 
presented in the main text, to obtain that the scattering regime 
is important when $q_t\gtrsim 20$. 
(At this time the fraction of 
energy in the fluctuations is about $5q_t^{-1/2}$.)
This leads to the initial value
$q_0\gtrsim (\Phi_0/M_{\rm P})^2(\ln g^{-1}/ 
\langle \mu_k\rangle)^2\sim 10^5$.
\bibitem{I(b)} The fact that scatterings of $\chi$ particles off the
$\phi$ zero mode are not important in this case can be verified by
looking at the $\phi$ field occupation numbers. We find that $n_k^\phi
< 300$ at all times and for all $k$ during our simulation. Similarly, 
$\langle (\delta\phi)^2\rangle \lesssim 4\times 10^{-11}\Phi_0^2$
at all times, which may be compared to figure 2.
\bibitem{chi-energy}
One can estimate the energy in the $\chi$ 
fluctuations at the time when the slowly varying state begins
as follows: the scatterings via the $\lambda_\chi\chi^4$ term
become fast when $n_{k_{\rm res}}^\chi\sim 10^2/\lambda_\chi$. The 
energy density at that time is 
$\rho_\chi\sim (n_{k_{\rm res}}^\chi/2\pi^2)k_{\rm res}^4\sim
g\Phi_0^2\omega_\phi^2/\lambda_\chi$, so that the fraction of 
the total energy in $\chi$ fluctuations is about 
$\rho_\chi/(\rho_\phi)_0\sim g/\lambda_\chi$.
\bibitem{static vs expanding} {This is true only for $m_\chi=0$, but 
remains approximately true for $m_\chi$ sufficiently small. It also 
requires that the Universe is radiation dominated, as is indeed the 
case for models with pure $\lambda \phi^4$ potentials
\cite{ShtaTrashBran}. 
In this case $a(\tau)\propto\tau$ and $\tau\propto t^{1/2}$. 
}
\end{thebibliography}
\end{document}